\documentclass[acmlarge]{acmart}
\AtBeginDocument{%
  \providecommand\BibTeX{{%
    \normalfont B\kern-0.5em{\scshape i\kern-0.25em b}\kern-0.8em\TeX}}}

\setcopyright{acmlicensed}
\copyrightyear{2024}
\acmYear{2024}
\acmDOI{XXXXXXX.XXXXXXX}

\begin{document}

\title{A Survey on Hardware Accelerators for Large Language Models}

\author{Christoforos Kachris}
\email{kachris@uniwa.gr}
\affiliation{%
  \institution{University of West Attica}
  \streetaddress{P.O. Box 1212}
  \city{Athens}
  \country{Greece}
}

\begin{abstract}

Large Language Models (LLMs) have emerged as powerful tools for natural language processing tasks, revolutionizing the field with their ability to understand and generate human-like text. As the demand for more sophisticated LLMs continues to grow, there is a pressing need to address the computational challenges associated with their scale and complexity. This paper presents a comprehensive survey on hardware accelerators designed to enhance the performance and energy efficiency of Large Language Models. By examining a diverse range of accelerators, including GPUs, FPGAs, and custom-designed architectures, we explore the landscape of hardware solutions tailored to meet the unique computational demands of LLMs. The survey encompasses an in-depth analysis of architecture, performance metrics, and energy efficiency considerations, providing valuable insights for researchers, engineers, and decision-makers aiming to optimize the deployment of LLMs in real-world applications.

\end{abstract}



\received{2024}

\maketitle

\section{Introduction}

Large language models (LLMs), such as GPT-3, are a class of artificial intelligence systems designed to understand, generate, and process human language with a level of complexity and nuance previously unattainable. These models are trained on extensive datasets, encompassing a broad spectrum of human discourse, which enables them to perform a wide array of language-related tasks, from translation and summarization to conversation and content creation. Over the past few years, the field of LLMs has witnessed remarkable evolution, characterized by significant advancements in model architecture, training methodologies, and data processing capabilities. 

This evolution has been driven by breakthroughs in deep learning and the increasing availability of large-scale computational resources. The impact of LLMs on society and business is profound and multifaceted. They have revolutionized the way we interact with technology, introducing new efficiencies and capabilities in areas such as customer service, content generation, and decision support systems. Moreover, LLMs have become instrumental in driving innovation across various sectors, including healthcare, finance, and education, by providing enhanced capabilities for data analysis, pattern recognition, and predictive modeling. This transformative influence underscores the importance of exploring and understanding the underpinnings of these models, their operational mechanisms, and the implications of their widespread application in diverse domains.

Until now there is not any comprehensive survey on the hardware accelerators to speedup the most computational intensive tasks of Transformers. In \cite{survey_transformer}, a survey is presented a survey on the hardware acceleration of transformer networks for autonomous driving. The paper presents several efforts on the acceleration of tasks such as object detection, 3D segmentation and lane detection. 

In 2022, Huang et al. presented a survey on hardware acceleration for transformers \cite{hw_survey_huang}. The paper was mostly focused on the the transformer model compression algorithm based on the hardware accelerator and was limited mostly on FPGA-based implementation. 

In 2023, Emani et al \cite{2023comprehensive} presented a comprehensive performance study of LLMs on several computing platforms and evaluated their performance characteristics for these models. They evaluated these systems using both micro-benchmarks, a GPT-2 model, and an LLM-driven science use case, called GenSLM. For the evaluation they use both GPUs, and other AI accelerators such as Sambanova, Cerebras, Graphcore and Habana Gaudi 2. Specifically they used LLM applications to evaluate the performance of the systems and categorize the proposed solution for the acceleration of the LLM applications.

In this paper we present a comprehensive survey on the several research efforts that have been presented for the acceleration of transformer networks for Large Language models and NLP using hardware accelerators. The survey presents the frameworks that have been proposed and then performs a qualitative and quantitative comparison regarding the technology, the processing platform (GPU, FPGA, ASIC), the speedup and the energy efficiency of each framework.

\section{Computational and energy requirements}

Computational complexity 

Large language models (LLMs) are among the most computationally intensive applications in contemporary artificial intelligence research. The computational intensity of LLMs is primarily attributed to their architecture and the scale of data they process. These models typically consist of hundreds of millions to billions of parameters, which are the learned weights that determine how the model processes and generates language. Training these parameters requires iterating over vast datasets multiple times, a process that demands substantial computational resources.

Moreover, the complexity of the tasks LLMs are designed to perform exacerbates their computational demands. Tasks such as contextual understanding, nuanced language generation, and handling ambiguities in human language require deep neural networks with multiple layers, each contributing to the overall computational load. The iterative nature of model training and fine-tuning, involving continuous forward and backward propagation through these deep networks, further amplifies the computational requirements.

\subsection{Computational Complexity in Training}

The training phase of Large Language Models (LLMs) is inherently complex due to the extensive computational resources required to process and learn from massive datasets. This complexity stems from several key factors:

\begin{itemize}
\item \textbf{Volume of Parameters:} LLMs, such as GPT-3, comprise an extraordinarily high number of parameters (often in the billions), each requiring computation during the training process. The adjustment of these parameters to minimize error involves extensive computational efforts.

\item \textbf{Size of Training Data:} LLMs are trained on vast datasets, necessitating significant computational power to process and learn from the data. This process includes tasks like tokenization, encoding, and the application of complex algorithms to understand linguistic patterns and structures.

\item \textbf{Depth of Neural Networks:} The deep neural network architectures employed in LLMs, consisting of multiple layers, add to the computational load. Each layer contributes to the model's ability to understand and generate language, requiring substantial computation for both forward and backward propagation during training.
\end{itemize}

\subsection{Computational Complexity in Inference}

While it is often presumed that the computational complexity of LLMs is predominantly a concern during the training phase, the inference phase also demands significant computational resources:

\begin{itemize}

\item \textbf{Model Size and Memory Footprint:} The large number of parameters in LLMs translates to a substantial memory footprint even during inference. Managing and accessing these parameters efficiently requires considerable computational resources, especially when processing complex queries or large volumes of data.

\item \textbf{Real-Time Processing Requirements:} For applications requiring real-time responses, such as chatbots or interactive tools, the LLM must process input and generate output rapidly. This necessitates the allocation of significant computational resources to ensure low latency and high throughput.

\item \textbf{Contextual and Nuanced Language Processing:} The ability of LLMs to understand context and nuance in language, although less computationally intensive than training, still requires a considerable amount of processing power. This is particularly true for tasks involving long context windows or complex linguistic structures.
\end{itemize}

In conclusion, both the training and inference stages of LLMs are characterized by substantial computational complexity. While the nature and scale of this complexity may differ between the two phases, both necessitate significant computational resources to achieve the high levels of performance and accuracy expected from state-of-the-art language models. This underlines the importance of ongoing research into more efficient model architectures and computational strategies to optimize both the training and application of LLMs.

\subsection{Energy consumption}

The immense computational requirements of LLMs translate into significant energy consumption. Training a state-of-the-art LLM can consume as much energy as the lifetime consumption of multiple cars, highlighting the environmental impact of these models. This energy consumption primarily stems from the use of high-performance GPUs or TPUs that are necessary for the parallel processing capabilities required in training these models.

The energy-intensive nature of LLMs raises concerns regarding their carbon footprint, especially when trained and operated in data centers powered by non-renewable energy sources. The continuous operation of these models for inference tasks also contributes to ongoing energy usage, although it is typically less than that required for training. The trade-off between the benefits offered by LLMs and their environmental impact is an ongoing area of research and debate, emphasizing the need for more energy-efficient model architectures and the adoption of green computing practices in the AI field.

Efforts to mitigate the energy consumption of LLMs include optimizing model architecture for efficiency, implementing more energy-efficient hardware, and utilizing renewable energy sources for data centers. These measures aim to reduce the environmental impact while maintaining the advanced capabilities of LLMs.



\section{FPGA-based accelerators}

\subsection{MNNFast}

In 2019, Jang et al. proposed MNNFast to accelerate large language models \cite{MNNFast}. MNNFast is accelerating large language models using three optimization. First, to reduce the memory bandwidth consumption, MNNFast proposes a new column-based algorithm with streaming which minimizes the size of data spills and hides most of the off-chip memory accessing overhead. Second, to decrease the high
computational overhead, MNNFast proposes a zero-skipping optimization to bypass a large amount of output computation. Lastly, MNNFast proposes an embedding cache dedicated to efficiently cache the embedding matrix 

MNNFast has been implemented across various platforms (CPU, GPU, and FPGA). MnnFast improves the overall throughput by up to 5.38×, 4.34×, and 2.01× on CPU, GPU, and FPGA respectively. The implementation that has been ported to FPGA achieves 6.54× higher energy efficiency. 


\subsection{FTRANS}

In 2020, Li et al \cite{2020_ftrans} presented a hardware acceleration framework, called FTRANS, that was targeting the acceleration of transformer-based large scale language representations. The proposed framework includes enhanced block-circulant matrix (BCM)-based weight representation to enable model compression on large-scale language representations at the algorithm level  which significantly reduces the model size (up to 16 times) with minimal accuracy degradation. 

FTRANS partitions the model into embedding layer and encoder/decoder stacks, with a focus on off-loading the embedding layer to off-chip memory to optimize computation. Additionally, it includes the development of a design automation and optimization technique for maximum throughput. FTRANS addresses the large size and complex data flow of transformer-based models by proposing a two-stage optimization approach to effectively schedule computation resources, optimizing latency and throughput on FPGA. 

Based on the performance evaluation, FTRANS achieves 27x speedup and 81x improvement in energy efficiency compared to CPU, and up to 8.8x improvement in energy efficiency compared to GPU.

\subsection{Multi-Head Attention}

In 2020, Lu et al. presented an FPGA based architecture for the acceleration of the most computationally intensive parts of transformer networks \cite{2020_multihead}. In their work they propose a novel hardware accelerator for two key components, i.e., the multi-head attention (MHA) ResBlock and the position-wise feed-forward network (FFN) ResBlock, which are the two most complex layers in the Transformer. 

Firstly, an efficient method is introduced to partition the huge matrices in the Transformer, allowing the two ResBlocks to share most of the hardware resources. Secondly, the computation flow is well designed to ensure the high hardware utilization of the systolic array, which is the biggest module the design. 
Thirdly, complicated nonlinear functions are highly optimized to further reduce the hardware complexity and also the latency of the entire system. 

The proposed framework is implemented on a Xilinx FPGA. Based on the performance evaluation the proposed design achieves a speed-up of 14.6× compared to a V100 GPU.

\subsection{FPGA NPE}

In 2021, Khan et al. presented an FPGA acceleration for language models called NPE. \cite{2021_FPGA_NPE}. The NPE architecture consists of an instruction control unit (ICU), a memory read unit (MRU), a memory write unit (MWU), a matrix multiply unit (MMU), and a nonlinear vector unit (NVU).

The MMU computation consists of five stages: data selection, inner product, adder tree reduction, accumulation, and quantization. Data selection loads the necessary matrix operands from the input buffers and rearranges them as needed. The matrix multiplication is implemented using an array of PEs, where each PE performs an inner product using an array of multipliers followed by an adder tree. The MMU implementation has 128 PEs with 16 multiply accumulate
units each (for a total of 2048 multipliers). These multiply accumulate units map to the FPGA’s DSP slices in our implementation.

The key novel component of NPE’s design is the nonlinear vector unit (NVU) that handles high-throughput nonlinear function computation with minimal resource overhead. The NVU is a data-parallel vector load/store architecture that performs arithmetic and logical operations on multiple elements in each clock cycle. 

NPE was implemented on Xilinx Zynq Z-7100 FPGA board clocked at 200 MHz. NPE is compared with other frameworks like FTRANS and implementation on CPU and GPU. Although that there is not any significant speedup compared to other computing platforms, the main advantage is the energy efficiency. NPE achieves around 4× better energy efficiency over CPU (i7-8700k) and 6× over GPU (RTX 5000). 

\subsection{Column Balanced Block Pruning}

In 2021, Peng et al. presented a novel scheme on accelerating Transformer networks using column balanced block-wise pruning \cite{2021_pruning}. The column balanced block-wise pruning combines the key features of both bank balanced pruning and block-wise pruning. The column balanced block-wise pruning ranks the blocks’ L2 norm by each column to get the pruning thresholds and prunes blocks for each column. 

By combining the Compressed Sparse Banks (CSB) format and the  Block Compressed Sparse Row (BCSR) format, a Compressed Sparse Column Block (CSCB) is formed.
A specialized process element (PE) is then introduced for the sparse matrix multiplication accelerator, and multiple PEs can be used to increase the accelerator throughput.

The proposed framework has been implemented on different hardware platforms (Intel i5-5257U (2.7 GHZ) CPU, Nvidia Jetson TX2 GPU, and Xilinx Alveo U200 FPGA) for further comparison of latency and throughput. The experimental results showed that the FPGA platform achieves a 11× speed up compared to the CPU platform and 2× speed up compared to the GPU platform.

\subsection{FPGA DFX}

In 2022, Hong et al. presented DFX \cite{2022_DFX} for the acceleration of the transformer networks used in LLMs. Similarly to NPE, the DFX architecture proposed a modular architecture consisting for several computer core for the acceleration of the transformer networks. 

To address the sequential characteristic of text generation, DFX compute core is optimized for single token processing. It also uses an efficient tiling scheme and dataflow architecture based on utilizing the high-bandwidth of HBM memory. DFX uses model parallelism on the multi-FPGA system to increase the physical number of compute cores that work in parallel while evenly assigning full workload to each device. DFX propose custom instruction set architecture (ISA) at the assembly language level to support the end-toend processing for inference. 
Similarly to NPE, the DFX core has two processing units, matrix processing unit (MPU) and vector processing unit (VPU). 

For the evaluation, DFX has been implemented on an Intel Xeon Gold 6226R CPU with four Xilinx Alveo U280 data center acceleration cards. DFX achieves an average of 3.8x throughput and 4x higher energy efficiency compared to the GPU appliances.

\subsection{FPGA OPU}

In 2023, Bai et al. proposed another scheme for the acceleration of transformer networks called Overaly OPU \cite{2023_OPU}. To support the inference of diverse networks, they propose a configurable computation unit. Specifically, they propose 48 processing elements (PEs) that are configured for the acceleration of the transformer networks. The output stage of the adder tree can be switched during the inference process. That way, data from forward modules can flow through the computation unit in pre-defined connection state. The proposed scheme achieves 5x-15× speedup compared with a CPU, 1.1-2.9× speedup compared with GPU (RTX 3090) and, 1.10-2.5× speedup compared with the other FPGA accelerators such as NPE \cite{2021_FPGA_NPE}.

\subsection{FPGA acceleration of Transformer networks}


In 2022, Tzanos et al, presented a high performance hardware accelerator for the transformer networks \cite{tzanos}.  Transformer networks use a technique called attention. The attention, adopted by the field of neuroscience, is the ability to be able to selectively concentrate on specific data while ignoring other data of the environment. In deep learning we imitate this technique through attention mechanisms and one way to achieve this is to encode a sequence not into a single fixed vector but to create a model that produces a vector for each output step by adding a set of weights which will later be optimized. 

Consequently it does not simply learn what to produce in the output but how to put weights selectively on specific input data maximizing the probability of a correct output. BERT is based using multiple attention blocks. Each
attention block converts the input using GEMM (General Matrix Multiplications) operations and then uses both GEMM and Non-Linear functions such as Softmax, Layer Normalization and Gaussian Error Linear Unit (GELU) to produce the output. 

The paper presents a novel architecture targeting FPGA platform that is used to speedup the specific functions; General Matrix Multiplications, Softmax, layer normalization and GELU. 

Since GEMM consumes about 70\% percent of the total execution time, the architecture was mostly focused on how to optimally accelerate the matrix multiplications operations of the network. Furthremore, the goal of this scheme was to integrate matrix multiplication, which are the most computational intensive parts of the Transformer network, with functions of very little impact on the runtime, such as Gelu, Softmax and Layernorm, in order to prevent unnecessary data transfers from the host to the FPGA kernel and vice versa something that it would definitely cost to our implementation. The kernels were developed on Vivado HLS targeting the Alveo U200 FPGA card and clocked at 411MHz. 

The performance evaluation showed that the proposed framework can achieve 2.3x system speedup for the BERT model compared to a 40-thread processor and 80.5x speed-up over a single-core CPU. 

\subsection{FlexRun}

In 2023, Hur at al. presented an FPGA-based accelerator to speedup the diverse and complex NLP models, called FlexRun \cite{hw_fpga_flexrun}. The paper is focused on accelerating both Recurrent Neural Networks (RNNs) models such as SRNN or long short term memory (LSTM) and attention-based NLP models, such as Transformer, and GPT2. 

FlexRun consists of three main schemes, FlexRun:Architecture, FlexRun:Algorithm, and FlexRun:Automation.

The main advantage of FlexRun is that it exploits the high reconfigurability of FPGAs to dynamically adapt the architecture to the target model and its configuration. The base architecture of FlexRun alleviates the overhead of vector operations by adopting a deeply pipelined architecture. The architecture consists of parameterized pre-defined basic modules that can be configured to fit the input model and its configuration. 

Next they perform a design space exploration on the algorithms to find the optimal compute unit design by finding the best modules for the input models using the FlexRun:Algorithm framework. Finally, using the results of FlexRun:Algorithm, they find the best architecture for the specific models and implement the architecture on the FPGAs using  tool called FlexRun:Automation. 

Based on the profiling of NLP and LLM applications they identify that the most computationally intensive tasks are the general matrix multiplication and the vector processing. Hence they propose a modular architecture that consists of general matrix multiplication units (Gemv-unit) and Vector processing unit. Gemv-units composes of multiple SIMD arithmetic units and the vector unit executes vector operations with some additional
operators (i.e., reduction, exp, gelu) to support attention-based NLP models.

For evaluation, they compare FlexRun with Intel’s Brainwave-like architecture on a Stratix-10 GX FPGA and a Tesla V100 GPU with tensor cores enabled. Compared to the FPGA baseline, FlexRun achieves an average speedup of 1.59× on various configurations of BERT. For GPT2, FlexRun gets 1.31× average speedup. Next, when comparing to the GPU implementation, FlexRun improves the performance by 2.79× and 2.59× for BERT and GPT2, respectively.

\subsection{ODE-based acceleration}

In 2024, a hybrid approach was proposed for the acceleration of the transformer networks by Okubo et al\cite{hw_fpga_ode}. The proposed scheme uses ResNet as a backbone architecture and replaces a part of its convolution layers with an MHSA (Multi-Head Self-Attention) mechanism. Using this approach they manage to significantly reduce the parameter size of such models by using Neural
ODE (Ordinary Differential Equation) as a backbone architecture instead of ResNet. The proposed hybrid model reduces the parameter size by 94.6\% compared to the CNN-based ones without degrading the accuracy.

The proposed model was deployed on a modest-sized FPGA device for edge computing. applications To further reduce FPGA resource utilization, the researchers quantized the model following QAT (Quantization Aware Training) scheme instead of PTQ (Post Training Quantization) to suppress the accuracy loss. As a result, an extremely lightweight Transformer-based model can be implemented
on resource-limited FPGAs ideal for edge applications. 

The performance evaluation on a Xilinx Zynq UltraScale+ MPSoC platform shows that the proposed FPGA implementation achieves 12.8× speedup and 9.2× energy efficiency compared to an ARM Cortex-A53 CPU implementation.

\section{CPU and GPU-based Accelerators}

\subsection{SoftMax}

In 2022, Choi et al. presented a novel framework for acceleration of transformer networks through Recomposing Softmax Layers\cite{2022_softmax}. The softmax layer normalizes the elements of the
attention matrix to values between 0 and 1. This operation is conducted along the row vector of the attention matrix. Based on the profiling, the softmax layer in the scaled dot-product attention (SDA) block uses 36\%, 18\%, 40\%, and 42\% of the total execution time of BERT, GPT-Neo, BigBird, and Longformer, respectively.

In their work, they propose to decompose the softmax layer into sub-layers to match their data access patterns with adjacent layers. Then, by fusing the decomposed softmax sub-layers with the subsequent and preceding operations they manage to reduce the off-chip memory traffic and thus reducing the total execution time for this layer. 

Softmax recomposition achieves up to 1.25×, 1.12×, 1.57×, and 1.65× speedups in inferring BERT,
GPT-Neo, BigBird, and Longformer on a A100 GPU by significantly reducing the off-chip memory traffic.

\subsection{LightSeq2}

In 2022, Wang et al. proposed a series of GPU optimizations to accelerate the training for a general family of Transformer models on GPUs called LightSeq2 \cite{2022_lightseq2}. 

LightSeq2 proposes 3 techniques for the acceleration of the training of transformer networks. 
Firstly, to all types of transformers, LightSeq2 uses fused kernel operators for both encoder and decoder
layers. Adjacent fine-grained element-wise kernels are fused into one coarse-grained kernel,
resulting in fewer kernel launches and intermediate results. For example, the last kernel of the self-attention layer implements bias adding, dropout, and residual kernels with only one kernel launch.

LightSeq2 also applies Mixed-Precision Update for the training. During the initialization of the
trainer, LightSeq2 copies all pieces of parameters/gradients into one tensor. Then it resets and links them as fragments of workspace. During each training step, LightSeq2 only executes the trainer kernel once to update the workspace, which prevents launching huge amount of fragmented GPU kernels on every piece of parameters/gradients.

Finally, LightSeq2 reduces the memory footprint by compacting memory with fewer allocation-and-releases at no extra cost. LightSeq2 divides the GPU memory into permanent memory with a fixed size to store parameters and gradients, and temporary memory with variable sizes to store intermediate tensors.

The performance evaluation shows that LightSeq2 is consistently faster (1.4-3.5×) than previous systems on different GPUs and it can achieve up to 3x speedup on large public datasets.

\subsection{Simplified Transformer Networks}

He and Hofmann \cite{he2023simplifying} have also proposed a novel framework to accelerate transformer networks in GPUs by simplified transformers without compromising convergence properties and downstream task performance.

Based on signal propagation theory and empirical evidence, they find that many parts can be removed to simplify GPT-like decoder architectures as well as encoder-style BERT models. Specifically, they show that it is possible to remove skip connections, value parameters, projection parameters and sequential sub-blocks, all while matching the standard transformer in terms of training speed and downstream task performance.

Based on the performance evaluation both on autoregressive decoder-only and BERT encoder-only models, the simplified transformers emulate the per-update training speed and performance of standard transformers, while enjoying 15\% faster training throughput in GPUs, and using 15\% fewer parameters. 

\subsection{LLama}
In 2023, Microsoft presented LLMA \cite{2023_microsoft_llama}, an LLM accelerator that is used to speed up Large Language Model (LLM) inference with references. LLMA was mainly motivated by the observation that there are many identical text spans between the decoding result by an LLM and the reference that is available in many real world scenarios (e.g. retrieved documents). LLMA first selects a text span from the reference and then copies its tokens to the decoder. Then it checks the tokens’ appropriateness as the decoding result in parallel within one decoding step. The improved computational parallelism allows LLMA to achieve over 2x speed-up for LLMs with identical generation results as greedy decoding. However, this speedup is achieved only in use cases where significant overlap between in-context reference and outputs exists (e.g., search engines and multi-turn conversations).

The main advantage of LLMA compared to other efficient decoding algorithms such as Speculative Decoding \cite{xia2023speculative} and Speculative Sampling \cite{chen2023speculative} that need to introduce an additional efficient drafter model to generate a draft for checking, LLMA does not require an additional model and is much easier to implement and deploy. 

For the performance evaluation, they used the LLaMA model with 7B, 13B and 30B parameters and all the inferences are done in half floating numbers. For the 7B and 13B models, the inferences were performed in one NVidia 32G V100 GPU, and for the 30B model, the inference was performed on four NVidia 32G V100 GPUs on a single machine. The proposed scheme achieved from 2.19x up to 3.06x speedup depending on model sizes and scenarios.

\subsection{UltraFastBERT}

In 2023, Belcak and Wattenhofer presented a novel scheme for the acceleration of language models purely on software called UltraFastBERT \cite{2023_exponentially}. UltraFastBERT selectively engages just 12 out of 4095 neurons for each layer inference. This is achieved by replacing feedforward networks with fast feedforward networks. The intermediate layers of UltraFastBERT are exponentially faster by design: given a feedforward (FF) and a fast feedforward (FFF) network, each with n neurons, the time complexity of a forward pass through the FFF is O(log2 n) instead of O(n) as for FF. 

UltraFastBert achievies 78x speedup over the optimized baseline feed-forward implementation, and the PyTorch implementation delivers 40x speedup over the equivalent batched feed-forward inference.

\section{ASIC Accelerators}

\subsection{A3}

One of the early research on the acceleration of transformer networks was proposed in 2020 by Hma et al. called A3 \cite{ham2020a3}. The paper proposes a hardware accelerator for attention mechanisms in NNs, that not only focused
on the efficient implementation of the attention mechanism in hardware but also on reducing the amount of
computation in attention mechanism through algorithmic optimization and approximation. It presents an approximate candidate selection mechanism to reduce the number of search targets, and thus the amount of computation. 

Additionally they propose a specialized hardware pipeline exploiting parallelism to accelerate approximated attention mechanisms enhancing not only the performance of the system but also increasing the energy efficiency. 

The proposed scheme has not been implemented on FPGA but is has been implemented on a cycle-accurate Verilog design  targeting a TSMC 40nm ASIC clocked at 1GHz. Based on the performance evaluation, the proposed scheme can achieve up to 7x speedup compared to a Intel Gold 6128 CPU implementation and up to 11x better energy efficiency against versus a CPU implementation.

\subsection{ELSA}

In 2021, Ham et al. presented a hardware-software Co-design approach for the acceleration of transformer networks called Elsa\cite{2021_elsa}. 

Based on the fact that irrelevant relations can be effectively filtered out by computing approximate similarity,
ELSA substantially reduces computational waste in a selfattention operation. Unlike conventional hardware such as CPU or GPUs, that cannot benefit from approximation, ELSA propose a specialized hardware that directly translates this reduction to further improve performance and energy efficiency. 

The approximate self-attention scheme consists of three sub-operations. The first one, estimates the angle between two vectors (e.g., a key and a query) with minimal computation by utilizing the concise representations of the key and the query. Then, an estimated angle is utilized to compute the approximate similarity between a query and a key, based on the fact that dot product is directly proportional to the cosine of the angle between two vectors. Finally, the approximate similarity is compared with a certain threshold to identify whether a specific key is relevant to the query or not.

They evaluate several representative self-attention-oriented NN models to demonstrate the effectiveness of the ELSA. For performance evaluation, they implemented a custom simulator for ELSA targeting a 40nm ASIC clocked at 1GHz. ELSA-moderate achieves up to 157x speedup compared to GPUs and two orders of magnitude improvements in energy efficiency over the GPU for the self-attention computation. 

\subsection{SpAtten}

In 2021, Want et al. presented a framework for the acceleration for large language models called Spatten. \cite{2021_spatten}

SpAtten propose a novel scheme for the acceleration of NLP using three algorithmic optimizations: cascade token pruning, cascade head pruning and progressive quantization to reduce computation and memory access. 

Pruning is applied to the tokens and heads, and not to weights as it usually the case. Cascade means that once a token/head is pruned, it is removed in all following layers, so one layer only needs to process remaining tokens/heads from previous layers. The deeper the layer, the more tokens/heads are pruned. 

The three techniques are input-dependent since the pruned computation and bit-width are adaptive to input instances. Cascade pruning requires sorting token/head importance scores on the fly. 
The proposed hardware architecture is based on high parallelism of top-k engines for token/head selections, specialized memory hierarchy, and fully-pipelined datapath to improve the performance and reduce the energy consumption. 

The proposed scheme has been implemented on a cycle-accurate design using SpinalHDL and mapped to ASIC using a 40nm TSMC library. SpAtten  and achieves 162x, and 347x speedup over a GPU (TITAN Xp), and a Xeon CPU, respectively. In terms of energy efficiency SpAtten achieves 1193x and 4059x energy savings compared to GPU and CPU.

\subsection{Sanger}

Lu at el. presented in 2021 another novel approach for the acceleration of transformer networks called Sanger\cite{2021_sanger}. 

Sanger accelerates the sparse attention models by combining dynamic sparsity patterns and reconfigurable architecture.
The software part provides sparsity patterns, which can achieve high performance and a balanced workload. The architecture is designed with reconfigurability to support the dynamic characteristics of sparsity, which helps to improve the compression ratio. 

As software level Sanger propose an algorithm to predict the dynamic sparsity by computing a low-bit version of the attention matrix and zeroing out small attention weights via binary thresholding. To ensure load balance for efficient hardware implementation, Sanger encodes the resulting attention mask with unstructured sparsity into multiple fine-grained structured blocks. The prediction and encoding are performed on the fly as Sanger determines the sparsity patterns
dynamically on a per-sample basis.

At the hardware level, Sanger proposes a score-stationary dataflow that unifies the computation of SDDMM and SpMM operations. Using this dataflow, Sanger keeps the sparse scores stationary in the processing elements until the computation is finished, which effectively avoids the decoding overhead. 

To allow more flexibility in the sparsity patterns, Sanger proposed a reconfigurable systolic array based on this dataflow. Sanger was implemented in Chisel hardware that was translated to Verilog RTL. The design was targeting an ASIC using the UMC 55nm technology clocked at 500MHz.

\subsection{Energon}

In 2023, Zhou et al. presented a algorithm-architecture co-design approach that accelerates various transformers using dynamic sparse attention, called Energon \cite{2023_energon}. Energon proposes a mix-precision multi-round filtering (MP-MRF) algorithm to dynamically identify query-key pairs at runtime. 

Energon adopts low bit-width in each filtering round and only the finally selected pairs are used for high-precision tensors in the attention stage to reduce overall complexity. By this means, they manage to reduce by 4× to 8× the computation cost with negligible accuracy loss. 

Energon is implemented as a co-processor targeting a 45nm ASIC library. 
Based on the performance evaluation it is shown Energon achieves 168× and 8.7× speedup and up to 10000× and 1000× energy reduction over Intel Xeon 5220 CPU and NVIDIA V100 GPU, respectively.

\section{In-Memory Hardware Accelerators}

\subsection{ATT}

In 2020, Guo et al. presented another approach for the acceleration of attention-based accelerators called ATT \cite{2020_att} based on resistive RAM. ATT is based on crossbar-based resistive RAM that can eliminates weight movement between memory and processing units with a dedicated pipeline design for Attention-based Neural Networks. The proposed scheme consists of several modules. 

The QKV engine employs a crossbar-based ReRAM to perform the matrix-vector multiplications. The Mask issuer computes masks according to the matrix fetched by the Q-K-V engine. Two different masks are output from this module. One is forwarded to the Attention engine, and the other is stored in the Mask cache to filter inputs and outputs of the Fully Connected engine.

Finally, the attention engine is used to perfor the matrix-to-matric multiplications in threee-stages. The first stage is is used for the computing of the inner-product between q vectors and the transposition of k vectors. The second is to softmax the previous innerproduct results. The third is multiplying the softmax results by the v vectors. 

The proposed scheme has been simulated using CACTI 7.0 at 32 nm to model the power and area of the SRAM buffer and the Mask Cache. Based on the performance evaluation, ATT can achieve 202x speedup compared to NVIDIA GTX 1080 Ti GPU.

\subsection{ReTransformer}

In 2020, Yang et al. proposed an in-memory framework for the acceleration of transformers called ReTransformer \cite{hw_retransformer}. ReTransformer is a ReRAM-based In-Memory architecture for Transformer acceleration that is not only accelerate the scaled dot-product attention of Transformer using ReRAM-based In-Memory architecture but also eliminate some data dependency by avoiding writing the intermediate results using the proposed matrix decomposition technique. Furthermore, ReTransformer proposes a new sub-matrix pipeline design for multi-head self-attention. 

ReRAM is used for performing the Matrix Multiplications of the transformer networks and the scale operations. ReTransformer is also optimizing the SoftMax by incorporating in-memory logic to maximally exploit on-chip ReRAM
crossbar arrays in the processing sub-arrays. Specifically, ReTransformer is using ReRAM-based compare and select logic to find the Maximum and use look-up tables to perform the exponential and logarithm functions. 

The performance evaluation shows that compared to GPU, ReTransformer can achieve up to 23.21× speedup while the corresponding overall power is reduced by 1086×.

\subsection{iMCAT}

In 2021, Laguna et al. presented a novel in-memory architecture for the acceleration of transformer networks for long sentences called iMCAT \cite{hw_inmemory}. The proposed framework uses a combination of XBars and CAMs to accelerate transformer networks. The acceleration of transformer networks is achieved by combining several techniques such as computing in-memory, thus minimizing the memory transfer overhead, caching reusable parameters to reduce the number of operations, exploiting the available parallelism in the attention mechanism, and finally using locality sensitive hashing to filter the number of sequence elements by their importance. 

The performance evaluation shows that this approach achieves a 200x speedup and 41x energy improvement for a sequence length of 4098.

\subsection{X-Former}

In 2023, Sridharan et al. presented a novel in-memory hardware acceleration to speedup transformer networks called X-Former\cite{hw_xformer}. X-Former is a hybrid spatial in-memory hardware accelerator that consists of both NVM and CMOS processing elements to execute transformer workloads efficiently. 

X-Former is composed primarily of a Projection Engine with NVM processing tiles for executing MV MStatic operations and an Attention Engine with CMOS processing tiles for executing MV MDynamic operations. The main difference compared to other in-memory architectures is that the weights of all the layers are stored in the Projection Engine to prevent reprogramming the NVM tiles, while the Attention Engine is optimized to only process the largest self-attention layer due to area constraints. 

X-Former proposes a sequence blocking dataflow, which overlaps the computations of the two processing elements and reduces the total execution time. X-Former has also dedicated bus-based interconnect network for data transfers between the Projection engine and the Attention engine that is used to reduce the energy consumption and the latency. 
Compared to the previous in memory schemes, X-Former avoids the frequent reprogramming by mapping static operations to NVM crossbars while processing the dynamic operations using in-memory CMOS processing elements.

Based on the performance evaluation it is shown that X-Former achieves up to 85x and 7.5x improvements in latency and energy over a NVIDIA GeForce GTX 1060 GPU and upto 10.7x and 4.6x improvements in latency and energy over a state-of-the-art in-memory NVM accelerator.

\subsection{Flash}

In 2024, Apple presented a novel scheme for the efficient deployment of transformer networks utilizing Flash memories\cite{hw_flash}. The paper tackles the challenge of efficiently running LLMs that exceed the available DRAM capacity by storing the model parameters in flash memory, but bringing them on demand to DRAM. The propose method constructs an inference cost model that takes
into account the characteristics of flash memory, that is used to optimize in two critical areas:
reducing the volume of data transferred from flash and reading data in larger, more contiguous chunks. Within this hardware-aware framework, Apple introduces two principal techniques. 

The first technique, called “windowing”, reduces data transfer by reusing previously activated neurons, and second, “row-column
bundling”, tailored to the sequential data access strengths of flash memory, increases the size of data chunks read from flash memory. These methods collectively allows to run models up to twice the size of the available DRAM, with a 4-5x and 20-25x increase in inference speed compared to naive loading approaches in CPU and GPU, respectively.

\section{Quantitative comparison}

The following table shows all of the hardware-based accelerators that have been proposed and the main features for each accelerator. Each row presents the name of the accelerator, the type of the accelerator (FPGA/GPU/ASIC/In-memory), the speedup, the energy-efficiency and the reference platform for the comparison. In some cases, the proposed scheme is compared against both a CPU and a GPU. In these cases, both speedup and energy efficiencies are shown. In this section we present a quantitative and qualitative comparison of the proposed schemes. 

\begin{table*}
  \caption{LLM-Transformer Accelerators}
  \label{tab:commands}
  \begin{tabular}{lllccc}
    \toprule
Year & Framework                     & Technology    & Speedup   & Energy efficiency & Baseline \\
    \midrule

2019 & MNNFast\cite{MNNFast}         & FPGA 7020     & 5.4x      &  6.5x   & CPU (Xeon 24-core)\\
2020 & FTRANS\cite{2020_ftrans}      & FPGA VCU118   & 27x-81x   &  8.8x   & CPU/GPU RTX5000  \\
2020 & Multi-Head\cite{2020_multihead} & FPGA XCVUP13 & 14x      & ---     & GPU V100\\
2021 & NPE\cite{2021_FPGA_NPE}       & FPGA Zynq7100 & 35x       &  4x-6x  & CPU/GPU RTX5000 \\
2021 & Pruning\cite {2021_pruning}   & FPGA U200     & 11x/2x    & ---     & CPU(i5)/GPU TX2\\
2022 & DFX\cite{2022_DFX}            & FPGA U280     & 3.8x      & 4x      & GPU V100   \\
2022 & Transformer\cite{tzanos}      & FPGA U200     & 2.3x      & ---     & CPU (80-threads) \\
2023 & OPU\cite{2023_OPU}            & FPGA          & 15x/2.9x  & ---     & CPU(Gold)/GPU RTX3090\\
2023 & FlexRun\cite{hw_fpga_flexrun} & FPGA S10      & 2.7x      & ---     & GPU V100\\
2024 & ODE\cite{hw_fpga_ode}         & FPGA          & 12.8x     & 9.2x    & ARM A53\\              

2022 & SoftMax\cite{2022_softmax}    & GPU A100      & 2.5x      & ---     & GPU A100\\
2022 & LightSeq2\cite{2022_lightseq2} & GPU A100     & 3x        & ---     & GPU A100\\
2023 & LLMA \cite{2023_microsoft_llama} & GPU V100   & 2x        & ---     & GPU V100\\
2023 & UltraFastBERT\cite{2023_exponentially} & CPU  & 78x       & ---     & CPU \\

2020 & A3\cite{ham2020a3}            & ASIC 40nm     & 7x        & 11x     & CPU(Gold)/GPU Volta\\
2021 & ELSA\cite{2021_elsa}          & ASIC 40nm     & 157x      & 1265x   & GPU V100/TPU \\
2021 & SpAtten\cite{2021_spatten}    & ASIC 40nm     & 347x/162x & 4059x/1093x & CPU(Xeon)/GPU Titan\\
2021 & Sanger\cite{2021_sanger}      & ASIC 55nm     & 22.7/4.64x & ---    & CPU(3970)/GPU V100 \\
2023 & Energon\cite{2023_energon}    & ASIC 45nm     & 168x/8.7x & 10000x/1000x & CPU/GPU V100\\

2020 & ATT\cite{2020_att}            & In-memory     & 202x      & 11x     & CPU(Gold)/GPU Volta  \\
2020 & ReTransformer\cite{hw_retransformer} & In-memory & 23x    & 1086x   & GPU \\
2022 & iMCAT\cite{hw_inmemory}       & In-memory     & 200x      & 41x     & GPU Titan RTX\\
2023 & X-Former\cite{hw_xformer}     & In-memory     & 85x       & 7.5x    & GPU GTX1060\\    
2023 & Flash\cite{hw_flash}          & Flash         & 25x/5x    & ---     & CPU/GPU\\    
    
    \bottomrule
  \end{tabular}
\end{table*}

\includegraphics[width=\textwidth]{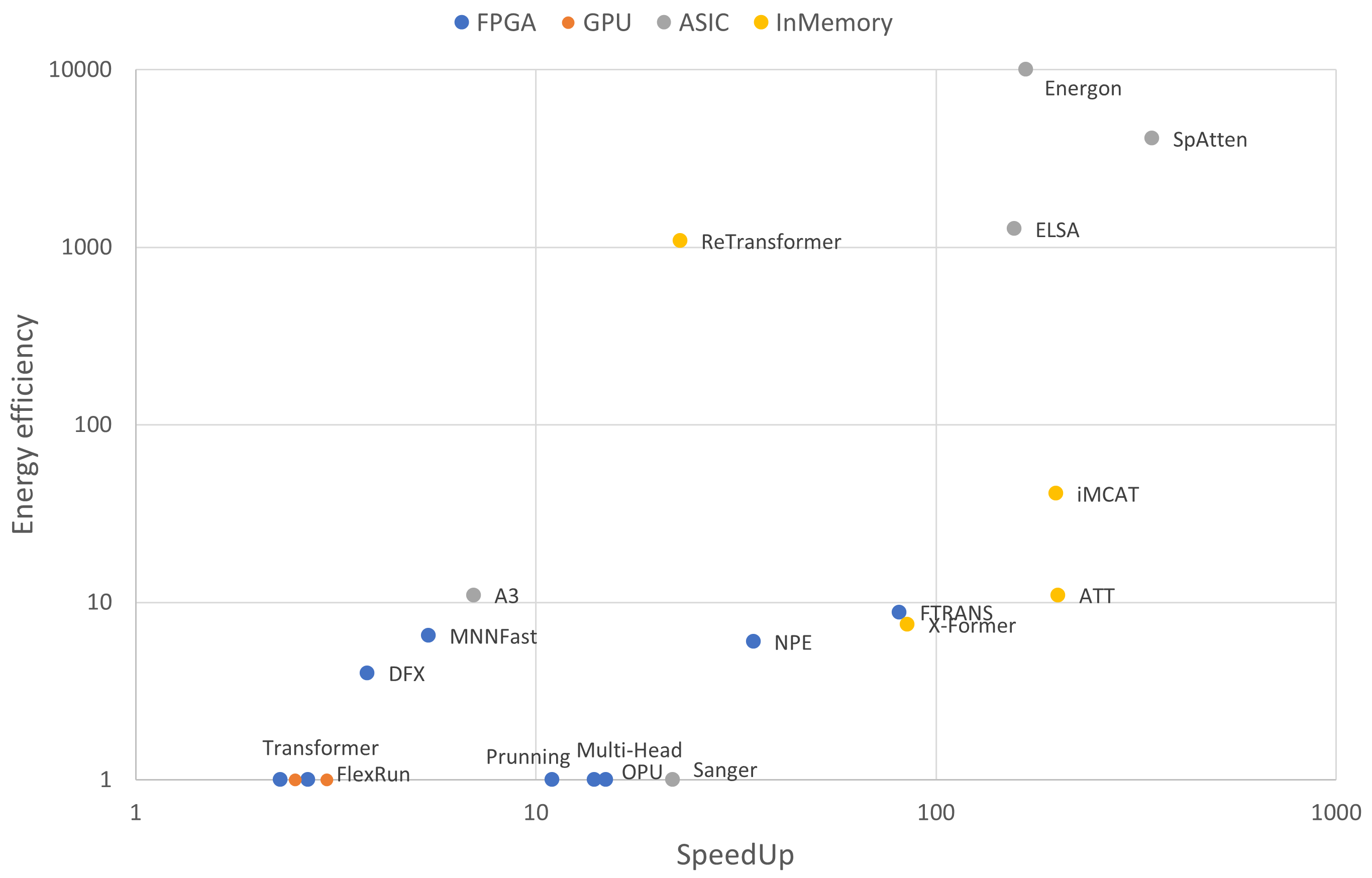}

\subsection{Technology}

There are four main computing platforms that researchers are utilizing to accelerate LLM and transformer networks; GPUs, FPGAs, ASIC and in-memory computing. GPUs is the most obvious platform as many LLMs are utilizing GPUs for the training and the inference. FPGAs can be customized with tailored-made architectures that are used to speedup specific functions. The main idea is to offload the most computationally intensive tasks from the CPU to the FPGAs. That way developers can keep the flexibility of the CPUs and at the same time to speedup the applications. The main advantage of the FPGAs is that the proposed scheme can be implemented and evaluated on the FPGAs in order to measure the real overall speedup. 

There are also several schemes that target an ASIC implementation. These schemes propose a custom architecture that can be used as a co-processor to speedup the most computationally intensive tasks like the matrix multiplications, etc. Although that none of the proposed schemes have been implemented in a real ASIC, all of the schemes have been implemented in a cycle-accurate design and have been evaluated using cycle-accurate simulators and power estimators like CACTI \cite{cacti}. Most of the schemes have been evaluated on a 40nm process technology while some of these schemes are implemented using 45nm or 55nm processes. 

Finally, some accelerators are targeting in-memory computing. In-memory computing is a promising technology that can utilize NVM and memristors to process data such as matrix multiplications. Although in-memory computing has many advantages is not yet widely used in the industry and it has several limitations in the commercialization. 

An interesting exception on this list is MNNFast \cite{MNNFast}, where the proposed optimizations have been applied on several platforms like CPU, GPU and FPGAs. In the case of CPU, they implemented MNNFast in C++ with open-source BLAS library, OpenBLAS. In the case of GPUs, they used  cuBLAS \cite{cublas} provided with CUDA Toolkit 10.0 to perform matrix-to-matrix multiplications. Finally, in the case of FPGAs, they designed and implemented an FPGA-based accelerator for MnnFast using Vivado High-Level Synthesis (HLS).

\subsection{Speedup}

The speedup is the execution time of the proposed scheme divided by the reference design. However, each accelerator is compared against a different reference computing platform making hard to evaluate the higher speedup against the same reference platform. Furthermore, some of the accelerators are using a reference platform based on CPU while other are using a reference design based on GPU. Even in the cases that a scheme is compared against a CPU it is not always clear if the reference design is a single-core CPU or a multi-core implementation on a CPU. However, there are some general conclusions that can be drawn based on the comparison. 

As it is shown in the table, ASIC and in-memory based accelerators tend to provide much higher speedup compared to FPGA or GPU-based accelerators. In-memory transformer accelerators can achieve up to 200x speedup while SpAtten achieves up to 347x speedup compared to CPUs. However, ASICs and in-memory computing need a huge investment in terms of time and money to fabricate it. FPGAs can provide lower speedup (up to 81x in the case of FTrans) but these platforms can be utilized immediately without the additional cost and time of the ASICs. 

\subsection{Energy efficiency}

The energy efficiency refers to the total energy consumption to perform an operation compared to the reference platform. Again, each scheme is compared against different computing platform making hard to evaluate the most energy efficient solution. However, similar to speedup, ASICs and In-memory computing accelerators provide much better energy efficiency compared to FPGAs and GPUs. Energon claims up to 4 order of magnitude better energy efficiency compared with a CPU and 3 orders of magnitude better energy efficiency compared to GPUs (V100). SpAtten and ELSA, both targeting a 40nm process technology achieve also 3 orders of magnitude better energy efficiency compared to GPUs utilizing the algorithmic optimizations. However, as in the case of speedup, FPGAs can provide lower energy efficiency but are readily available and can be integrated with off-the-shelf components in the current data centers. 

\section{Conclusions}

Large Language Models have been emerged as a promising and powerful technology for the science and the society in general. However, the huge amount of computational complexity pose a new challenge in data centers as these applications consume huge amounts of energy. Hardware accelerators can be used to speedup these applications and reduce significantly the energy requirements. The architectures that have been proposed so far show that hardware accelerators can be customized to speedup the most demanding functions of the transformer networks and can be used to reduce the energy requirements in the data centers by more than 4 orders of magnitude. The reduction in the energy consumption results also to lower carbon emissions and lower water consumption for cooling affecting. As LLMs and LMMs will continue to increase in complexity and processing requirements the utilization of hardware accelerators will be necessary for the future data centers.


\bibliographystyle{ACM-Reference-Format}
\bibliography{sample-base}

\appendix

\end{document}